\documentstyle[osa,manuscript,psfig]{revtex}

\begin{document}       

\title{Spin alignment of the high $p_T$ vector mesons in
polarized $pp$ collisions at high energies}

\author {Xu Qing-hua and Liang Zuo-tang}
\address{Department of Physics,
Shandong University, Ji'nan, Shandong 250100, China}

\date{}
\maketitle

\begin{abstract}   
Using the same method as that used in $e^+e^-$ annihilation and
lepton-nucleon deep-inelastic scattering, 
we calculate the spin alignment of vector mesons with 
high transverse momenta 
in high energy $pp$ collisions with one longitudinally polarized beam.
We present the results obtained at RHIC energies.
We also study the spin alignment of such vector mesons
in the case of a transversely polarized proton beam
and present its relation to the transversity distribution in the nucleon.
Numerical results obtained using simple models for 
the transversity distributions are also given.
These results show that there exist significant spin alignment 
in both cases, and they can be measured in experiments 
such as those at RHIC.
\end{abstract} 

\newpage
\section{Introduction}
Spin effects in high energy fragmentation processes
have attracted much attention recently
[See, e.g., Refs. [1-12] and the references given therein].
One of the important issues is the study of 
spin transfer from the fragmenting quark to the produced hadrons.
It can be studied by measuring the polarizations of the hadrons
produced in the fragmentation of a polarized quark.  
Hyperon polarization has been frequently used in this 
connection\cite{Chen95,Jaf96,Ada97,BL98,Kot98,Ansel01,Mas,LL2000,XLL02}.
On the other hand, it has been pointed out\cite{XLL01,XL02} that
the vector meson polarization can also be used for this purpose
since the polarization of a vector meson can also be studied  
easily in experiments. 
Furthermore, compared with hyperon,
the production rate of a vector meson is usually higher 
and the origin is simpler
due to the much smaller contribution from the decay of heavier hadrons.
This implies less theoretical uncertainties in calculations and 
better statistics in experiments. 
Therefore, the study of polarization of the vector mesons may provide us
important information to understand the role of spin
in hadronic reactions. 

The polarization of a vector meson is
described by the spin density matrix $\rho$ or its element $\rho_{m,m'}$,
where $m$ and $m'$ label the spin component along the quantization axis, 
usually the $z$-axis of the frame. 
The diagonal elements $\rho_{11}$, $\rho_{00}$ and $\rho_{-1-1}$
for the unit-trace matrix are the relative intensities of meson spin
component $m$ to take the values $1$, $0$, and $-1$ respectively.
In experiment,
$\rho_{00}$ can be measured from the angular
distribution of its decay products and
a deviation of $\rho_{00}$ from $1/3$ indicates spin alignment.
In the helicity basis, i.e., in the case that the
$z$-axis is chosen as the moving 
direction of the vector meson, the matrix is usually called
helicity density matrix and 
$\rho_{00}$ represents the probability
of the vector meson to be in the helicity zero state.
Measurements have been carried out in different
reactions\cite{epr82,BEBC87,EXCHARM00,Kress01} in particular
in $e^+e^-$ annihilation at LEP recently \cite{Kress01}.
The data\cite{Kress01} show that 
$\rho_{00}$ is significantly larger than
$1/3$ for vector mesons with large $z$ and increases with increasing
$z$ (here $z\equiv2p_V/\sqrt{s}$, where $p_V$ is the momentum of
the vector meson, $\sqrt{s}$ is the total $e^+e^-$ center of mass energy).
This means that there exist a significant spin alignment for vector
mesons in helicity frame in $e^+e^-$ annihilation at the $Z^0$ pole,
and the effect is more significant for the large momentum fraction region.
The simple statistic model\cite{Bigi} based on spin counting gives the 
result of $\rho_{00}=1/3$.
A QCD-inspired model is given in Ref. [\ref{Augustin80}], 
where a fast quark combines
a soft anti-quark to form a vector meson and the soft anti-quark
preferentially has the same helicity as the fast one.
This model leads to $\rho_{00}=0$.
Another model\cite{Donoghue} describes the production
of vector meson through the channel $q\to qV$, with the 
vector meson coupling to the quark like a vector current.
In this case, the helicity-conserving vector current
gives rise to $\rho_{00}=1$. 
None of these predictions is consistent with the data\cite{Kress01}.

In a recent paper\cite{XLL01}, we calculated
the helicity density matrix of vector
mesons inclusively produced in  $e^+e^-$ annihilation 
at $Z^0$ pole by taking the spin of a vector meson
which contains a fragmenting quark as the
sum of the spin of the polarized fragmenting
quark and that of the anti-quark created in the fragmentation process.
By comparing the obtained results with the data\cite{Kress01}, 
we showed that the experimental results for $\rho_{00}$  
imply a significant polarization for the anti-quark
which is created in the fragmentation process and combines with the
fragmenting quark to form the vector meson.
It should be polarized in the opposite direction
as that of the fragmenting quark and
the polarization can approximately be written as,
\begin{equation}                        
P_z = -\alpha P_f ,
\label{eq1}
\end{equation}
where $\alpha \approx 0.5$ is a constant;
$P_z$ is the polarization of the anti-quark in the moving direction
of the fragmenting quark and $P_f$ is the longitudinal polarization
of the fragmenting quark of flavor $f$.
The results of $\rho_{00}$ obtained from above relation
are in good agreement with the data for different mesons. 
Since the calculations are rather straight-forward,
the relation given by Eq. (\ref{eq1}) can be considered as a
direct implication of the data\cite{Kress01}
in $e^+e^-$ annihilation.
It implies that there exists a spin interaction between
the fragmenting quark and the anti-quark 
during the hadronization process.
It should be interesting to extend the calculations 
to other processes and make further tests to
see whether the relation is universal, in the sense that
it is true for quark fragmentation in all different reactions.
Hence we made similar calculations for polarized lepton-nucleon
deep-inelastic scattering (DIS) process and polarized $pp$ collisions.
The results for DIS have been given in Ref.[\ref{XL02}].
In this paper, we present the calculations and the results 
for high $p_T$ vector mesons in polarized $pp$ collisions, 
where factorization theorem can be used so that 
fragmentation effects 
and other effects such as those from the structure functions 
can be studied separately.  

Compared with those in the lepton-induced reactions
such as $e^+e^-$ annihilations and DIS,
the study of high $p_T$ vector mesons in $pp$ collisions
has the following properties. 
(1) The involved hard scatterings are  
strong interaction processes rather than
the electroweak processes in the lepton-induced reactions,
so the corresponding cross section should in general be larger.
Furthermore, the luminosity of the incoming proton beams
can in general be made higher than that for leptons.   
Hence, the statistics in experiment can be better.
(2) Not only longitudinally but also transversely polarized
quarks can be produced, so that we can study the properties 
in both cases.  
(3) There are many different hard subprocesses which contribute
to the high $p_T$ hadron production.
This makes the study more interesting and,
at the same time, more complicated
than those in lepton-induced reactions.  
For example, the gluon fragmentation is also involved here,
which is unclear even for the unpolarized case.
It is very important to know whether its contribution is
significant in different kinematic regions.
We will first make an estimation of it in next section
using a Monte Carlo event generator and then  
present the calculation method of spin alignment 
of vector mesons in $pp$ collisions. 
In section III, we calculate the spin alignment of vector mesons
with high $p_T$ in the helicity frame
in $pp$ collisions with one longitudinally polarized proton beam
and present the results.
In Section IV, we extend the study to the
the spin alignment of high $p_T$ vector mesons  
for the case that one proton beam is transversely polarized.
Finally, a short summary is given in Section V. 
\section{The calculation method}
In this section, we first summarize the calculation method
of spin alignment of vector mesons that was
used in $e^+e^-$ annihilation, and then extend it to the 
inclusive high $p_T$ vector meson production
in longitudinally polarized $pp$ collisions.
\subsection{For $e^+e^-$ annihilation process}
The calculation method for vector mesons produced
in $e^+e^-$ annihilation has been presented in Ref.[\ref{XLL01}].
The main points are summarized as follows.
For $e^+e^-$ annihilation at a given energy, 
we need to consider vector mesons produced in the fragmentation
of a polarized initial quark $q^0_f$ of flavor $f$ and those of
polarized initial anti-quark ${\bar q}^0_f$.
The polarization of $q^0_f$ or ${\bar q}^0_f$ is a 
given constant which can be calculated using the 
standard model for electroweak interaction. 
To calculate the spin density matrices of these vector mesons,
we divide them into the following two groups
and consider them separately:
(A) those which contain the fragmenting quark $q^0_f$;
(B) those which don't contain the fragmenting quark.
The spin density matrix $\rho^V(z)$ for the vector meson $V$
is given by:          
\begin{equation}
\rho^V(z)=
\sum_f
\frac {\langle n(z|A,f)\rangle}{\langle n(z)\rangle}
\rho^V(A,f)
+\frac {\langle n(z|B)\rangle}{\langle n(z)\rangle}
\rho^V(B)
,
\label{eq2}
\end{equation}
where $\langle n(z|A,f)\rangle$ and $\rho^V(A,f)$
are respectively the average number and spin density matrix 
of the vector mesons from group (A);
$\langle n(z|B)\rangle$ and $\rho^V(B)$ are those from (B);
$\langle n(z)\rangle$=
$\sum_f \langle n(z|A,f)\rangle$+
$\langle n(z|B)\rangle$ is the
total number of vector mesons
and $z$ is the momentum fraction of the initial quark's momentum
carried off by the vector meson.
The average numbers $\langle n(z|A,f)\rangle$
and $\langle n(z|B)\rangle$ are independent of the spin properties
and can be calculated using an
event generator based on a fragmentation model 
which gives a good description of the unpolarized data.
We used generator $\sc pythia$\cite{pythia} in our calculations.

In terms of the usually used cross section and fragmentation
functions, Eq.(\ref{eq2}) has the following form,  
\begin{equation}
\rho^V(z)=
\frac{\sum_f D_{V/f}^A(z)\rho^V(A,f)
\sigma_{e^+e^-\to q^0_f{\bar q}^0_f}}
{\sum_f D_{V/f}(z)\sigma_{e^+e^-\to q^0_f{\bar q}^0_f}}
+\frac{\sum_f D_{V/f}^B(z)\rho^V(B)
\sigma_{e^+e^-\to q^0_f{\bar q}^0_f}}
{\sum_f D_{V/f}(z)\sigma_{e^+e^-\to q^0_f{\bar q}^0_f}}    
,
\label{eq3}
\end{equation}   
where $D_{V/f}^A(z)$ and $D_{V/f}^B(z)$ are the fragmentation
functions of the initial quark $q_f^0$ (anti-quark) into
the vector mesons of group (A) and (B) with momentum fraction $z$ respectively;
$D_{V/f}(z)=D_{V/f}^A(z)+D_{V/f}^B(z)$ is the usual 
fragmentation function;
$\sigma_{e^+e^-\to q^0_f{\bar q}^0_f}$ is the cross section
of the process $e^+e^-\to q^0_f{\bar q}^0_f$.
We see that,
$D^A_{V/f}(z)\sigma_{e^+e^-\to q^0_f{\bar q}^0_f}$
is just the cross section of the production of the meson $V$
which contains the initial quark via  
the fragmentation of $f$-flavor quark in $e^+e^-$ annihilation,
i.e., $e^+e^-\to q^0_f{\bar q}^0_f \to V(q_f^0 \bar q)X$.
It is proportional to
the average number $\langle n(z|A,f)\rangle$
in Eq. (\ref{eq2}).
The denominator is the total cross section of inclusive 
vector meson production, which corresponds to 
$\langle n(z)\rangle$ of Eq. (\ref{eq2}).

There are many different possibilities for the production
of of the vector mesons in group (B), 
we take them as unpolarized, i.e., $\rho^V(B)$=1/3.
For those vector mesons from group (A),
i.e., those which contain $q^0_f$ and
an anti-quark $\bar q$ created in the fragmentation,
the spin is taken as the sum of the spins of
$q_f^0$ and $\bar q$.
The polarization of $q_f^0$ is taken as the same 
as that before the fragmentation.
Then the spin density matrix $\rho^V(A,f)$ can be calculated
from the direct product of the spin density matrix
$\rho^{q^0_f}$ for $q^0_f$
and  $\rho^{\bar{q}}$ for $\bar{q}$.
Transforming the direct product,
$\rho^{q^0_f{\bar q}}$=$\rho^{q^0_f}$$\otimes$$\rho^{\bar q}$,
to the coupled basis
$|s, s_z\rangle$ (where $\vec{s}$=${\vec {s}^q}$+${\vec {s}^{\bar q}}$),
we can obtain the spin density matrix of the vector meson $\rho^V(A,f)$.
In the helicity frame of $q_f^0$, i.e., $z$-axis is taken as
the moving direction of $q_f^0$,
the density matrix of the vector meson of type (A) is given by, 
\begin{equation}
\rho^V(A,f)=\frac{1}{3+P_fP_z}
\small {
\left (
\begin{array}{ccc}
(1+P_f)(1+P_z)                   &
 \frac{{1+P_f}}{\sqrt2}(P_x-iP_y)                &
 0                               \\
\frac{(1+P_f)}{2}(P_x+iP_y) &
 (1-P_fP_z)&
 \frac{{1-P_f}}{\sqrt2}(P_x-iP_y) \\
0                               &
 \frac{1-P_f}{\sqrt2}(P_x+iP_y)                &
 (1-P_f)(1-P_z)                   \\
\end{array}  \right)
,}
\label{eq31}
\end{equation}            
where $P_f$ is the longitudinal polarizations of $q^0_f$ and
$\vec P=(P_x,P_y,P_z)$ is the polarization vector of 
the anti-quark $\bar q$.
Hence, the 00-component of the density matrix takes 
the following simple form,
\begin{equation}
\rho_{00}^V(A,f)=(1-P_fP_z)/(3+P_fP_z),
\label{eq4}
\end{equation}                   
in which there is only one unknown variable $P_z$, 
the polarization of $\bar q$ along the 
moving or polarization direction of $q_f^0$.

Using Eqs. (\ref{eq2}) and (\ref{eq4}), we can
determine $P_z$ in different cases by fitting 
the data\cite{Kress01} in the $e^+e^-$ annihilation at $Z^0$ pole 
for the production of different vector mesons.
As mentioned in the introduction, we found that\cite{XLL01},
the data for different vector mesons can reasonably be fitted 
if we take the $P_z$ satisfying the relation given by Eq.(1).
Now, we insert Eq.(1) into Eq. (\ref{eq4}) and obtain that\cite{note1}, 
\begin{equation}
\rho_{00}^V(A,f)=
{(1+\alpha P_f^2)}/{(3-\alpha P_f^2)}.
\label{eq5}
\end{equation}
Finally, from Eqs. (\ref{eq2}) and (\ref{eq5}), we have
\begin{equation}
\rho_{00}^V(z)=\sum_f
\frac{1+\alpha P_f^2}{3-\alpha P_f^2}
\frac {\langle n(z|A,f)\rangle}{\langle n(z)\rangle}
+ \frac{1}{3}
\frac {\langle n(z|B)\rangle}{\langle n(z)\rangle}
.
\label{eq6}
\end{equation}    
\subsection{Polarization of the outgoing quark in the hard 
subprocess in $pp$ collisions}
\begin{figure}[h]
\psfig{file=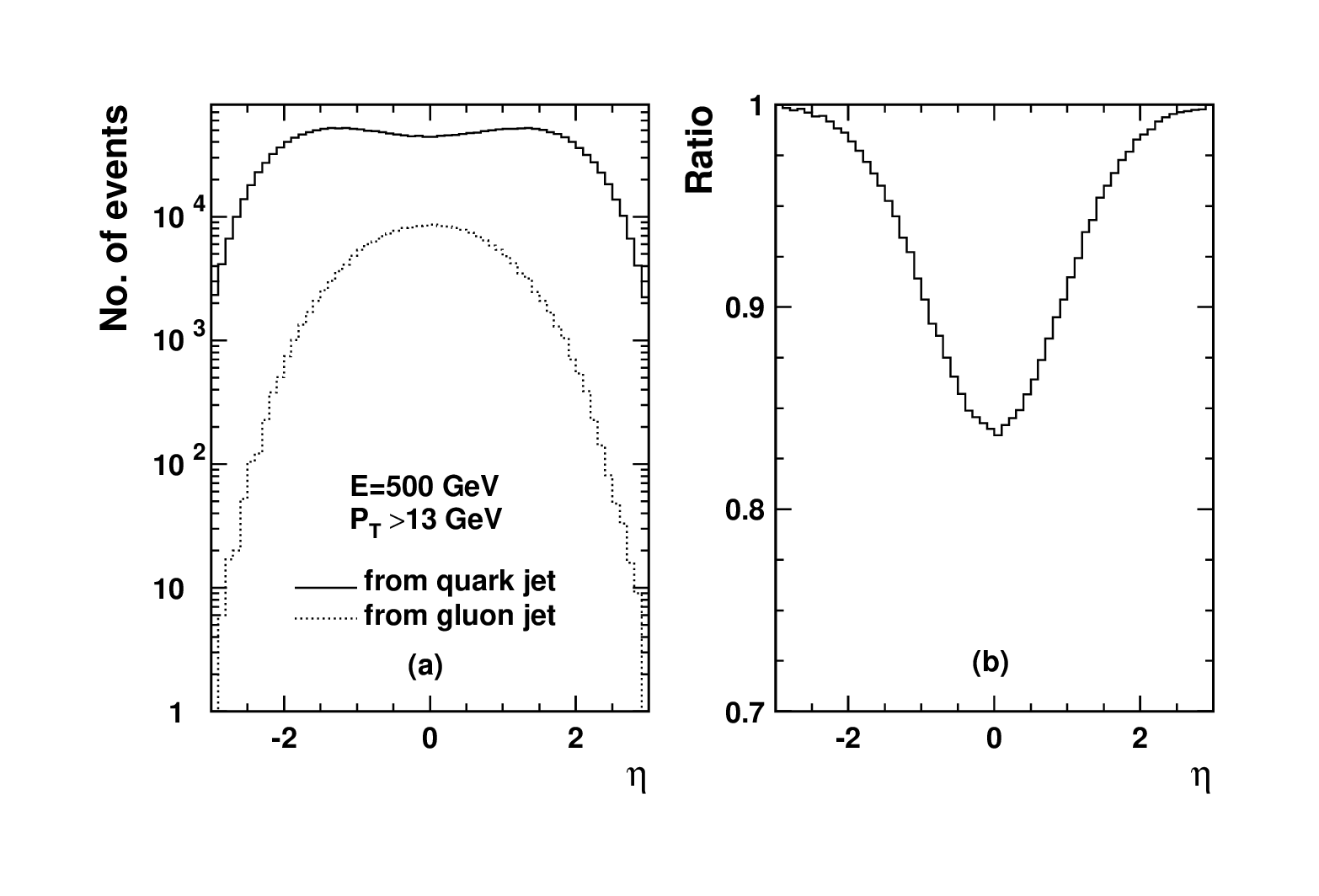,width=15cm}
\caption{(a) Contributions of quark and gluon fragmentation 
to $\rho^{0}$ production, and (b) the ratio of the quark fragmentation's
contribution to the total rate,
in $p p$$\to$$\rho^{0}X$ at $\sqrt s$=500 GeV and $p_T$$>$13 GeV.}
\label{meta132}
\end{figure}

Now we come to the high $p_T$ vector meson production in 
polarized $pp$ collisions.
Such vector mesons come mainly 
from the fragmentation of outgoing partons in the hard
scattering subprocesses.
The outgoing parton can be a gluon or a quark, and
the cross section of gluon-involved subprocess is 
even larger than others.
However,
the gluon distribution and fragmentation functions are both poorly
known yet, especially for the polarized case. 
Fortunately, the momentum fractions carried by the gluons
in a proton are usually very small, so the gluon contributions
to very high $p_T$ hard scattering subprocesses are suppressed.
In addition, it is known that the gluon fragmentation function
is softer than the quark fragmentation function. 
Consequently, for the final hadron production with high $p_T$, 
the contribution from gluon fragmentation is much smaller
than that from quark fragmentation\cite{XLL02}.
To see it numerically, we calculate these contributions 
using $\sc pythia$.
In Fig. \ref{meta132}, we show the results for inclusive 
$\rho^0$ production with $p_T$$>$13 GeV
at $\sqrt s$=500 GeV in $pp$ collisions.
We see that, at such high $p_T$ cutoff, the contribution of
gluon fragmentation is indeed much smaller than that of quark 
fragmentation, especially for the large $\eta$ region
($\eta$ is the pseudo-rapidity of the vector meson).
Thus, as we did in studying hyperon polarization 
in high $p_T$ jets\cite{XLL02},
we neglect the contribution of polarized gluon
fragmentation to the spin alignment of vector mesons,
and consider only the contribution of
polarized quark's fragmentation.
  
For quark fragmentation,
we can use the same method of calculating spin alignment
as that used in $e^+e^-$ annihilation.
To do it, we need to know the polarization of the outgoing quark
in the hard subprocess.
Such quark is also longitudinally polarized when one proton
beam is longitudinally polarized and its
polarization can be calculated in pQCD.
The obtained polarization can be written as,
\begin{equation}
P_q^{}(x_a,Q^2,y)=D_L(y)P_q^{in}(x_a,Q^2)
\label{polq7}
\end{equation}
where $D_L(y)$ is the longitudinal polarization transfer factor in the
scattering from the incoming parton to the outgoing parton;
$D_L(y)$ is only a function of $y$ which 
is defined as $y\equiv k_b\cdot (k_a-k_c)/k_a\cdot k_b$
for the subprocess $ab \to cd$, where
$k_a$, $k_b$, $k_c$ and $k_d$ are the four-momenta
of the partons $a$, $b$, $c$ and $d$ respectively.
$D_L(y)$ has been calculated using pQCD and 
the leading order results can be found in many publications,
for example, in Ref.[\ref{XLL02}].
We take the parton $a$ as the parton 
from the polarized proton beam,
and the polarization of the incoming parton $a$ is given by
$P_q^{in}(x_a,Q^2)=\Delta q(x_a,Q^2)/q(x_a,Q^2)$, where
$\Delta q(x_a,Q^2)$ and $q(x_a,Q^2)$ are the helicity and unpolarized
distribution functions at momentum fraction $x_a$ and
scale $Q^2$.
Therefore, the longitudinal polarization $P_q^{}$ is
the function of variables $x_a$, $Q^2$ and $y$.
Since we are now discussing $pp$ collisions with one polarized beam,
it is clear that
the outgoing quarks with positive momenta in $z$ direction
have a much larger probability of being polarized than those with
negative momenta when the positive axis of $z$
is taken as the direction of the incoming polarized proton.  

\subsection{Calculation method for spin alignment in
longitudinally polarized $pp$ collisions}
Similar to that in $e^+e^-$ annihilation,
to calculate the spin density matrix of the high $p_T$ vector
mesons in $pp$ collisions, 
we also divide them into two groups (A) and (B)
according to whether they contain an outgoing quark in 
the hard scattering 
and consider them separately.
The difference is that, now the fragmenting quark is an outgoing
quark in the hard scattering subprocesses and its
polarization is not a constant. 
It depends on the origin of this quark and is given by Eq.(\ref{polq7}).
Obviously, it is a function of $x_a$, $Q^2$ and $y$.
Hence, $\rho_{00}$ for vector mesons of group (A) should
be also a function of $x_a$, $Q^2$ and $y$ and it is given by,
\begin{equation}
\rho_{00}^V(x_a,Q^2,y|A)=
{(1+\alpha P_q^2)}/{(3-\alpha P_q^2)}.
\label{eq8}
\end{equation}  
 
To obtain $\rho_{00}$ for vector mesons at a given 
pseudo-rapidity $\eta$,
we need to replace the products in Eq.(\ref{eq3}) with 
the corresponding convolutions,
i.e.,
\begin{equation}
\rho^V(\eta)=
\frac{\sum_f [D_{V/f}^A(z_c)\otimes\rho^V(A,f)+
D_{V/f}^B(z_c)\otimes\rho^V(B)]\otimes
\sigma_{pp\to q_fX}}
{\sum_f D_{V/f}(z_c)\otimes\sigma_{pp\to q_fX}}
\label{eqpp01}
\end{equation}  
To write it more precisely, we have,
\begin{equation}
\rho_{00}^V(\eta)=
\small{
\frac{
\int d^2{p_T}\sum_{abcd}dx_a dx_b
 f_a(x_a,\mu^2)f_b(x_b,\mu^2)
\frac {d \hat {\sigma}} {d\eta}
[D^A_{V/c}(z_c,\mu^2)\rho_{00}^V(A,c)+
D^B_{V/c}(z_c,\mu^2)\rho_{00}^V(B)]
}
{\int 
d^2{p_{T}}\sum_{abcd}
\int dx_a dx_b f_a(x_a,\mu^2)f_b(x_b,\mu^2)
\frac {d \hat {\sigma}} {d\eta}(ab \to cd)
D_{V/c}(z_c,\mu^2)
}},
\label{rhopp}
\end{equation}
where $p_T$ is the transverse momentum of the vector meson;
$f_a(x_a,\mu^2)$ and $f_b(x_b,\mu^2)$ are the unpolarized
distribution functions of partons $a$ and $b$ in proton
at the scale $\mu$,
$x_a$ and $x_b$ are the corresponding momentum fractions carried
by $a$ and $b$;
$D_{V/c}(z_c,\mu^2)$ is the usual fragmentation function of parton $c$
into vector meson $V$, $D^A_{V/c}(z_c,\mu^2)$ and
$D^B_{V/c}(z_c,\mu^2)$ are those of parton $c$ into
vector mesons of group (A) and (B) respectively,
$z_c$ is the momentum fraction of parton $c$ carried by the
produced $V$;
${d\hat{\sigma}}/{d\eta}$ is  the
cross section at the parton level.
The cross section of the hard subprocess
can be calculated using perturbative QCD.
The summation in Eq. (\ref{rhopp})
runs over all possible subprocesses.   

The fragmentation functions are independent of the spin properties
and can be calculated using an available Monte Carlo event generator.
In practice, we generate a $pp$ collision event and then
search for the produced vector meson $V$.
We then calculate its contribution to $\rho^V_{00}(\eta)$
by tracing back its origin:
If the vector meson $V$ belongs to group (B), then its
contribution to $\rho^V_{00}(\eta)$ is given by $\rho_{00}(B)=1/3$.
If it belongs to group (A), we then need to further trace back 
the origin of the fragmenting quark to calculate its polarization
$P_q$ using Eq.(\ref{polq7}). 
We insert the $P_q$ into Eq.(\ref{eq8}) to get the contribution 
$\rho_{00}(A)$
of such vector meson to $\rho^V_{00}(\eta)$.
After running the program for sufficiently large number of events,
we obtain the final $\rho^V_{00}(\eta)$.
In terms of a mathematic formula, the calculation in this
procedure is expressed as,
\begin{equation}
\rho_{00}^V(\eta)=
 [\sum_{i=1}^{ N(\eta|A)}\rho_{00}^i(x_a,Q^2,y|A)
+\frac{1}{3} N(\eta|B)]/
{ N(\eta)},
\label{eq11}
\end{equation}
where $ N(\eta|A)$ and $N(\eta|B)$ are the numbers 
of vector mesons of group (A) and (B) as a function of $\eta$,
and ${N(\eta)}=N(\eta|A)+ N(\eta|B)$
is the total number of vector mesons.
The results for $\rho^V_{00}(\eta)$ in longitudinally
polarized $pp$ collisions
are presented in next section.
\section{Results of longitudinally polarized case}
\begin{figure}[t]
\psfig{file=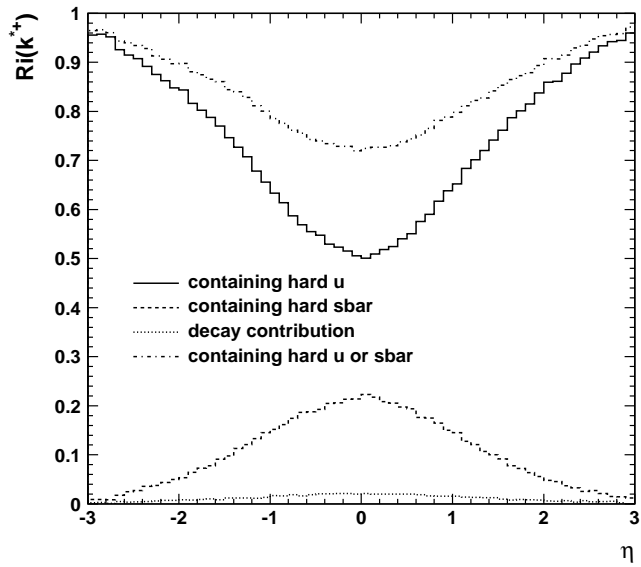,width=10cm}
\caption{Different contributions to $K^{*+}$ production
in $p p$$\to$$K^{*+}X$ at $\sqrt s$=500 GeV and $p_T$$>$13 GeV
as functions of $\eta$. Here ``containing hard u" and similar denote
the contributions of those containing the outgoing quark
from the hard scattering subprocess.}
\label{korg}
\end{figure}

\begin{figure}[t]
\psfig{file=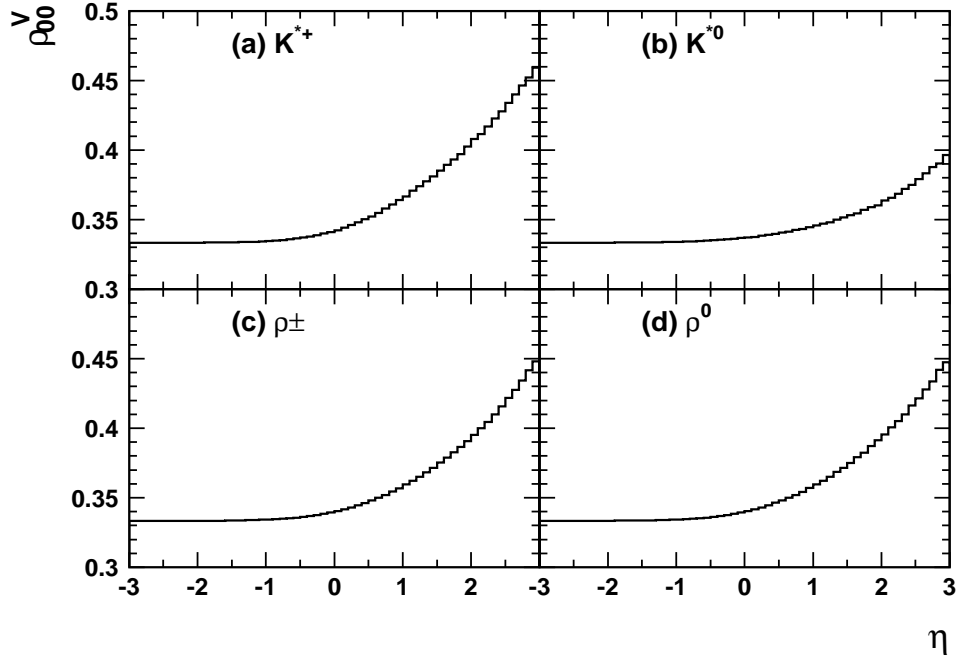,width=15cm}
\caption{Spin alignment of different vector mesons with $p_T$$>$13 GeV
in $p p$$\to$$VX$ at $\sqrt s$=500 GeV in helicity frame
when one proton beam is longitudinally polarized.
The standard LO set of GRSV2000\cite{GRSV2000}
and LO set of GRV98\cite{GRV98} are used for helicity 
and unpolarized distribution functions respectively.
The scale $\mu$ is taken as the transverse momentum of the
hard scattering subprocess.}
\label{vfig13}
\end{figure}

Using the method described in last section,
we calculate the $\rho_{00}$'s for 
different vector mesons in $pp$ collisions with one proton beam
longitudinally polarized. 
We now summarize the obtained results in the following.
Using the generator $\sc pythia$\cite{pythia},
we first calculate the different contributions to vector meson
production, i.e., those of group (A) and (B).
As an example, we show the results for $K^{*+}$ 
at $\sqrt s$=500 GeV and $p_T$$>$13 GeV 
in Fig. \ref{korg}.
From these results, we can see clearly that,
the decay contribution is indeed very small. 
The contribution of 
those containing an outgoing quark in the hard scattering
subprocess is very high, even larger than $90\%$ for large $\eta$ region.
The results of $\rho_{00}$'s for different vector mesons
are shown in Fig. \ref{vfig13}.
We see that, the spin alignment for $K^{*+}$, $\rho^{\pm}$ and $\rho^0$
are significantly high.
The obtained $\rho_{00}$'s for these mesons
increase from 1/3 to about 0.45 with increasing $\eta$.
There is nearly no spin alignment in the negative
$\eta$ region since the scattered quark moving in the negative
$z$ direction is almost unpolarized.
The spin alignment for $K^{*0}$ is smaller in both
positive and negative $\eta$ region,
this is because $d$ quark fragmentation dominate $K^{*0}$
production at high $p_T$ and the polarization of $d$ quark 
$|\Delta d(x,Q^2)/d(x,Q^2)|$ is
smaller than that of $u$ quark $|\Delta u(x,Q^2)/u(x,Q^2)|$.   

We can also make calculations for other vector mesons, 
such as $K^{*-}$, $\bar{K}^{*0}$, and $\phi$.
However, 
since these mesons do not contain $u$ or $d$ valence quark,
which dominate the high $p_T$ hadron production in $pp$ collisions,
their production rates at high $p_T$ are small.
This makes the statistics much worse than those for 
$K^{*+}$, $K^{*0}$, $\rho^{\pm}$ and $\rho^0$.
For the same reason, the $s$ quark distribution is much more
important. Hence,
the spin alignment for these mesons are much more sensitive to
the helicity distribution of the strange quark in nucleon, 
which is less precisely determined yet. 
It is thus not a good choice to study the spin transfer in
fragmentation process by measuring the spin alignment of these mesons.
But, it is possible to use them to study the helicity 
distribution of the strange sea in nucleon if they can be
measured with good accuracy.

\section{Transversely polarized case}

When one of the proton beams is transversely polarized,
the incoming quark $a$ for the elementary hard scattering 
process can also be transversely polarized in the same direction.
Its polarization $P_{aT}$ is determined by the 
transversity distribution function\cite{Trans} $\delta q(x_a,Q^2)$
and the unpolarized distribution function $q(x_a,Q^2)$,
i.e., $P_{aT}=\delta q(x_a,Q^2)/q(x_a,Q^2)$.
There is no gluon transversity distribution 
at leading twist\cite{Trans},
thus the incoming gluon in the hard scattering 
is not transversely polarized,
which is different from the longitudinally polarized case.
The transverse polarization can also be transferred 
from the incoming quark to the outgoing quark in the hard scattering.
The polarization direction of the outgoing quark 
is also transverse to the moving direction of the quark but is 
in general different from that of the incoming quark.
Both the magnitude and the direction of the polarization of the 
outgoing quark can be calculated\cite{Collins94} using pQCD for 
the hard elementary process. 
The results are summarized in the following.

We recall that, for a quark $a$ with transverse polarization $P_{aT}$, 
the spin density matrix $\rho_a$ in the helicity basis is given by,
\begin{equation}
\rho_a^{(in)}=\frac{1}{2}
\left (
\begin{array}{cc}
1 & P_{aT}e^{-i\phi}\\
P_{aT}e^{i\phi} & 1 \\
\end{array}  \right),
\label{eqrho1}
\end{equation}
where $\phi$ is the angle between 
the polarization direction and the $x$ axis. 
The $xy$-plane is perpendicular to the moving direction 
of the quark, which is taken as the $z$-direction; 
and the $x$-axis is taken as 
the normal direction of the scattering plane. 
The spin density matrix of 
the outgoing quark in the helicity basis can be
obtained from $\rho_a^{(in)}$ given in Eq. (\ref{eqrho1}) 
and the helicity amplitudes of the scattering. 
The result is given by\cite{Collins94}, 
\begin{equation}
\rho_q=\frac{1}{2}
\left (
\begin{array}{cc}
1 & P_{aT}D_T(y)e^{-i\phi}\\
P_{aT}D_T(y)e^{i\phi} & 1 \\
\end{array}  \right),
\label{eqrho2}
\end{equation}
where $D_T(y)$ is a real function of $y$ defined 
in Section 2B for the scattering 
and is called the polarization transfer factor.  
$D_T(y)$ for different subprocesses have been calculated 
using pQCD and can be found, for example, in Ref.[\ref{Collins94}]. 

From Eq. (\ref{eqrho2}), we see that the outgoing quark is 
also transversely polarized. 
The magnitude of the the polarization is given by $D_T(y)P_{aT}$, i.e, 
\begin{equation}
P_{qT}(x_a,Q^2,y)=D_T(y) \cdot \delta q(x_a,Q^2)/q(x_a,Q^2).
\label{eqpqT}
\end{equation}
We see also that, the angle between the polarization direction 
and the $x$-axis remains the same before and after the scattering. 
More precisely, 
the polarization vector of the outgoing quark 
is in the plane transverse to the moving direction 
of the quark and the angle between 
the polarization direction and the normal of the scattering 
plane is the same as that for the incoming quark. 
In another word, the direction of transverse polarization
of the incoming and that of the outgoing quark are related 
to each other by a rotation around the normal of the scattering plane, 
which changes the moving direction of the quark from 
the incoming to the outgoing direction.
(C.f. Fig. 2 of Ref.[\ref{Collins94}]).
We note that the direction of the polarization of 
the outgoing quark depends not only on 
the scattering angle but also on the azimuthal direction. 
In contrast, the magnitude of the polarization depends 
only on the scattering angle. 
The dependence is given by the corresponding 
spin transfer function $D_T(y)$, 
which is only a function of $y$ defined above.

Having seen that the outgoing quarks are also transversely 
polarized if the incoming protons are transversely polarized, 
we now discuss the spin alignment of the vector mesons produced in 
the fragmentation of such quarks. 
We recall that, 
the relation given in Eq. (\ref{eq1}) is obtained for 
the fragmentation of the longitudinally
polarized quarks by fitting the $e^+e^-$ annihilation data\cite{Kress01}. 
It shows that the anti-quark which combines the fragmenting quark
to form the vector meson is polarized in the opposite direction
as that of the fragmenting quark.  
To test whether the relation is in general true for 
the fragmentation of a longitudinally polarized quark 
in any high energy reactions, we extended the 
calculations of the spin alignment of vector mesons 
to other lepton-induced reactions in [\ref{XL02}] 
and longitudinally polarized $pp$ collisions in last sections. 
Now we further assume that the relation is also true 
for the fragmentation of a quark polarized in any direction, 
i.e, we extend the relation to,
\begin{equation}
\vec P_{\bar q}=-\alpha \vec P_{q}.
\label{barqT}
\end{equation}
We use this relation to calculate the spin alignment of
the vector mesons in transversely polarized $pp$ collisions
so that this extension can be tested in future experiments. 
The calculations are carried out in a way similar to that 
in the longitudinally polarized case.
We now present the calculations and the results 
in the following two different frames.

(1) In the helicity frame:
In the helicity frame of the outgoing quark of the hard scattering, 
its spin density matrix is off-diagonal and is 
given by Eq. (\ref{eqrho2}). 
From the relation given by Eq. (\ref{barqT}), 
the spin density matrix of 
the anti-quark which combines with the fragmenting quark 
to form the vector meson is,
\begin{equation}
\rho_{\bar q}=\frac{1}{2}
\left (
\begin{array}{cc}
1 & -\alpha P_{qT}e^{-i\phi}\\
-\alpha P_{qT}e^{i\phi} & 1 \\
\end{array}  \right).
\label{eqrho3}
\end{equation} 
We build the direct product 
of $\rho_q$ given by Eq.(\ref{eqrho2}) 
with $\rho_{\bar q}$ given by Eq.(\ref{eqrho3}), 
transform it to the coupled basis, and obtain the 
spin density matrix of the vector meson of group (A) as,  
\begin{equation}
\rho^{(h)}(A)=\frac{1}{3-\alpha P^2_{qT}}
\small {
\left (
\begin{array}{ccc}
 1  & \frac{1}{\sqrt 2}P_{qT}(1-\alpha)e^{-i\phi}&
 -\alpha P^2_{qT}e^{-2i\phi}             \\
\frac{1}{\sqrt 2}P_{qT}(1-\alpha)e^{i\phi}&
1-\alpha P^2_{qT} & 
\frac{1}{\sqrt 2}P_{qT}(1-\alpha)e^{-i\phi} \\
-\alpha P^2_{qT}e^{2i\phi} &
\frac{1}{\sqrt 2}P_{qT}(1-\alpha)e^{i\phi}& 1 \\
\end{array}  \right),}
\label{eqrho4}
\end{equation} 
where the superscript ``$(h)$" denotes the helicity frame.
Hence, the $\rho^{(h)}_{00}(A)$ for such vector 
mesons in the helicity basis is given by, 
\begin{equation}
\rho_{00}^{(h)}(x_a,Q^2,y|A)=
(1-\alpha P^2_{qT})/(3-\alpha P^2_{qT}).
\label{eq14}     
\end{equation} 
We can see that, as can be expected, 
if we look only at the $00$-component of the spin density matrix, 
the result is independent of the azimuthal direction of 
the transverse polarization, i.e. independent of the angle $\phi$. 
It depends only on the magnitude of the 
transverse polarization $P_{qT}$, 
thus on the variables $x_a, Q^2$ and $y$.
We see also that 
$\rho_{00}^{(h)}(A)$
is smaller than $1/3$, which is different
from the results in the longitudinally polarized case.
This shows that, if the fragmenting quark is 
transversely polarized, the produced vector meson 
should have a larger probability to be in the 
helicity $\pm 1$ states. 
As long as there exists the contribution of 
the vector mesons of group (A),
the final results for $\rho^{(h)}_{00}(\eta)$ in this frame 
should also be smaller than $1/3$.
We know from Fig. 2 that the contribution of the group (A)
increase with increasing $\eta$,
so we expect a decreasing $\rho^{(h)}_{00}(\eta)$ 
as $\eta$ increases for vector mesons such as $K^{*+}$ and $\rho^\pm$. 
It is significantly smaller than $1/3$ for large $\eta$.
Hence, the measurements of $\rho^{(h)}_{00}(\eta)$ 
of the vector mesons in transversely polarized $pp$ collisions
should give a good check of 
the extension of the relation of 
Eq. (1) to the transversely polarized case.

(2) In the transversity frame:
To compare with the results in the helicity frame  
for the longitudinally polarized case, we also calculate 
the spin alignment of the vector mesons 
in the transversely polarized case in the 
``transversity frame''. 
In this frame, 
similar to the helicity frame for longitudinally polarized case,
we take the (transverse) polarization direction of 
the fragmenting quark as the quantization axis.
The spin density matrix of the outgoing quark and that 
of the anti-quark are both diagonal in this frame, 
and they are given by
\begin{equation}
\rho_q^{(t)}=\frac{1}{2}
\left (
\begin{array}{cc}
1+P_{qT} & 0 \\
0 & 1-P_{qT} \\
\end{array}  \right),
\label{eqrho5}
\end{equation}
\begin{equation}
\rho^{(t)}_{\bar q}=\frac{1}{2}
\left (
\begin{array}{cc}
1-\alpha P_{qT} & 0 \\
0 & 1+\alpha P_{qT} \\
\end{array}  \right),
\label{eqrho6}
\end{equation}
where the superscript ``$(t)$" denotes the transversity frame. 
We see that they have completely the same form as 
those for the corresponding quarks and anti-quarks 
in longitudinally polarized case in the helicity frame. 
The corresponding spin density matrix $\rho^{(t)}(A)$,  
which can be obtained from the direct product 
$\rho_q^{(t)}\otimes\rho^{(t)}_{\bar q}$, 
is also diagonal, and the result for the  
$00$-component $\rho_{00}^{(t)}(A)$ is given by,
\begin{equation}
\rho_{00}^{(t)}(x_a,Q^2,y|A)=
(1+\alpha P^2_{qT})/(3-\alpha P^2_{qT}).
\label{eq15}
\end{equation}
We note that, since $\rho^{(t)}(A)$ 
is diagonal in the transversity frame, 
there is a simple relation between 
$\rho_{00}^{(h)}(A)$ and $\rho_{00}^{(t)}(A)$. 
The relation can easily be obtained by using 
$\rho_{00}^{(h)}(A)=\langle h=0|\rho(A)|h=0\rangle 
=\sum_{m_T,m_T'}\langle h=0|m_T\rangle 
\langle m_T|\rho(A)|m_T'\rangle 
\langle m_T'|h=0\rangle $ and 
$\langle m_T|\rho(A)|m_T'\rangle =
\rho_{m_T,m_T'}^{(t)}=\rho_{m_T,m_T}^{(t)} \delta_{m_T,m_T'}$. 
(Here, $h$ denotes the helicity and $m_T, m_T'$ denote 
the spin component along the quantization axis in 
the transversity frame.)
It is given by, 
\begin{equation}
2\rho_{00}^{(h)}(x_a,Q^2,y|A)+\rho_{00}^{(t)}(x_a,Q^2,y|A)=1.
\end{equation}
We see that, $\rho_{00}^{(t)}(x_a,Q^2,y|A)$  
can also be obtained from this relation and the result for 
$\rho_{00}^{(h)}(x_a,Q^2,y|A)$ given by Eq. (\ref{eq14}). 
We also see that, 
the obtained $\rho_{00}^{(t)}(x_a,Q^2,y|A)$ 
given by Eq.(\ref{eq15}) is in general larger than $1/3$. 
This means that, the probability for the produced vector mesons 
to be in the spin states which are transverse to the moving direction 
are relatively smaller. 
We see that, although the values of $\rho_{00}^{(t)}(x_a,Q^2,y|A)$ 
and $\rho_{00}^{(h)}(x_a,Q^2,y|A)$  are different, 
they represent the same physical meaning. 

We should note that, to measure $\rho_{00}^{(t)}$ in experiments,
we need to determine the quantization axis in this frame. 
Since quantization axis is chosen as the polarization direction 
of the quark before the fragmentation, 
it can be different for different vector mesons. 
The polarization direction is determined by Eq.(\ref{eqrho2}).
In practice, 
we first find out the normal direction of the 
scattering plane and then
determine the polarization direction according 
to the rules described below Eq. (\ref{eqpqT}).
(C.f., also Fig.2 of Ref.[\ref{Collins94}]).
The moving direction of the fragmenting quark 
can be replaced by the thrust axis of the quark jet.

\begin{figure}[t]
\psfig{file=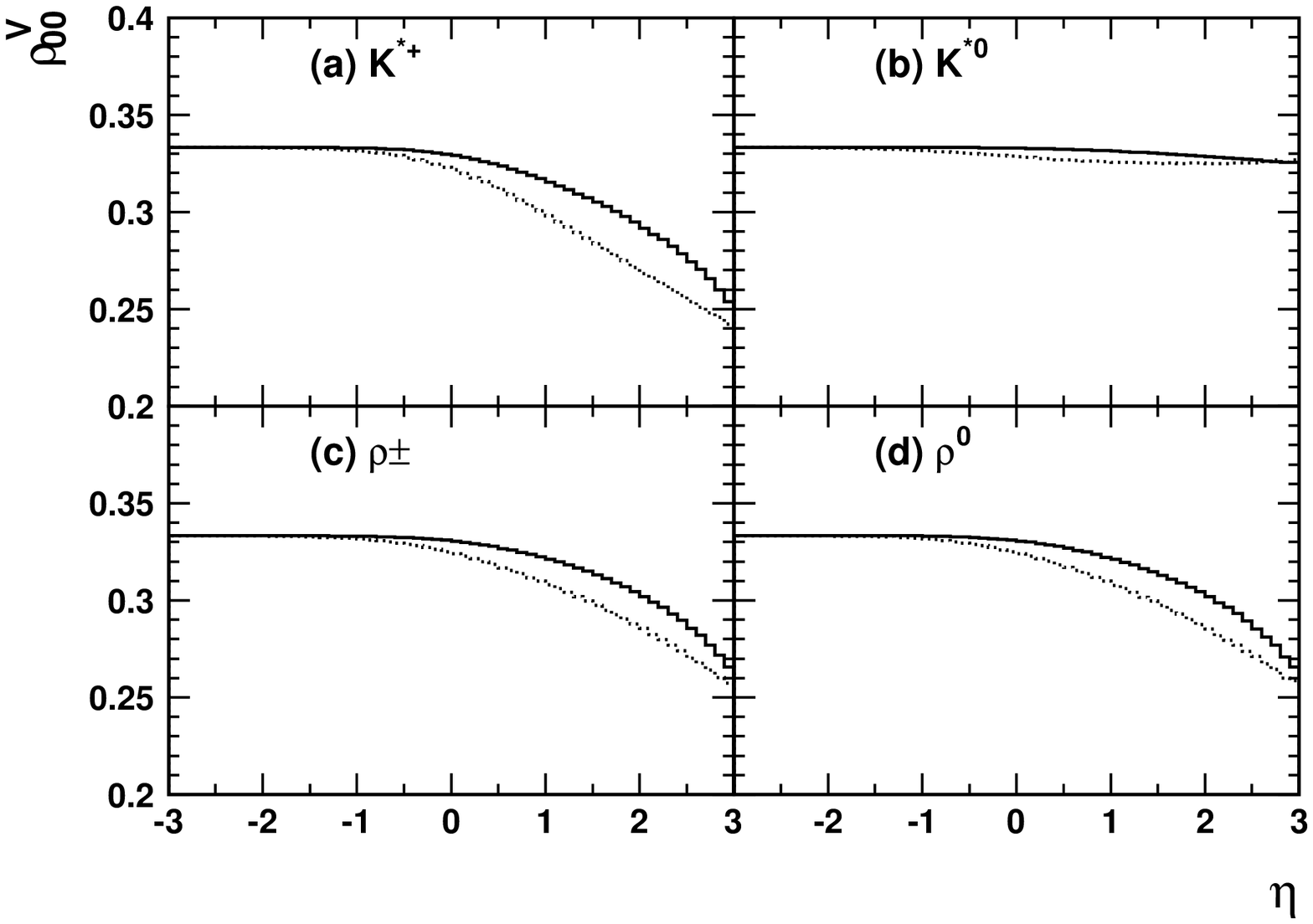,width=15cm}
\caption{Spin alignment of different
vector mesons with $p_T$$>$13 GeV
in $p p$$\to$$VX$ at $\sqrt s$=500 GeV in the helicity frame
when one of the proton beams is transversely polarized.
The solid lines denote the results obtained using the light-cone
quark-spectator model for $\delta q(x)$;
the dotted lines correspond to the results obtained using 
the upper limit in Soffer's inequality.}
\label{vfigt13h}
\end{figure}

\begin{figure}[t]
\psfig{file=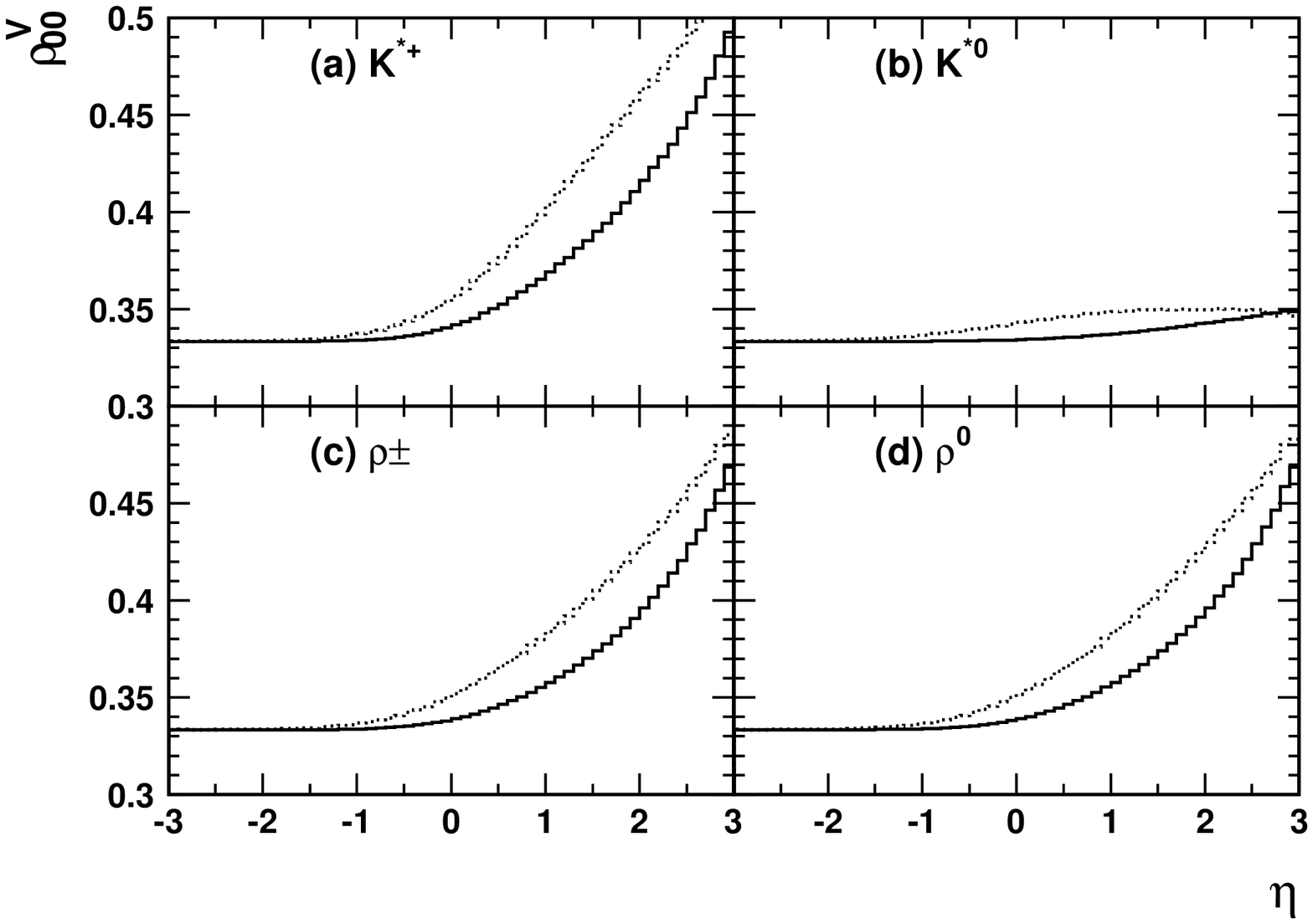,width=15cm}
\caption{Spin alignment of different 
vector mesons with $p_T$$>$13 GeV
in $p p$$\to$$VX$ at $\sqrt s$=500 GeV in the transversity frame
when one of the proton beams is transversely polarized.
The solid lines denote the results obtained using the light-cone
quark-spectator model for $\delta q(x)$;
the dotted lines correspond to the results obtained using 
the upper limit in Soffer's inequality.}
\label{vfigt13}
\end{figure}

Inserting Eqs. (\ref{eq14}) and (\ref{eq15}) into Eq.(\ref{eq11}),
we can calculate $\rho_{00}(\eta)$ for different vector mesons
in the above two frames 
in $pp$ collisions with one beam transversely polarized.
For these calculations, we first need to calculate $P_{qT}$
using Eq.(\ref{eqpqT}).
So far, the transversity distribution $\delta q(x,Q^2)$ is not known yet.
In order to get a feeling of how large the spin alignment can be,
we make an estimation using the simple form for $\delta q(x)$ 
in the light-cone SU(6) quark-spectator model\cite{BQMa} as input.
In this model, the $\delta q(x)$ is given by,
\begin{eqnarray*} 
\delta u_v(x)=[u_v(x)-\frac{1}{2}d_v(x)]{\widetilde {M}}_S(x)
-\frac{1}{6}d_v(x){\widetilde {M}}_V(x);
\end{eqnarray*}
\begin{equation}
\delta d_v(x)=-\frac{1}{3}d_v(x)\widetilde{M}_V(x),
\end{equation}
where $\widetilde{M}_S(x)$ and $\widetilde{M}_V (x)$ are related to 
the Melsh-Wigner rotation, which can be calculated according to the 
procedure given in [\ref{BQMa98}].
The obtained results of $\rho_{00}(\eta)$ for different vector mesons
at $\sqrt s=500$ GeV and $p_T>13$ GeV in the helicity frame
are shown in Fig. \ref{vfigt13h} and those in the transversity
frame are shown in Fig. \ref{vfigt13}.
For comparison, we also show the results obtained by
taking the upper limit of $\delta q(x,Q^2)$
in Soffer's inequality\cite{Soffer},
\begin{equation}
|\delta q(x,Q^2)| \le \frac{1}{2}[\Delta q(x,Q^2)+q(x,Q^2)].
\end{equation}
From these figures, 
we see that, in the helicity frame, the $\rho_{00}$'s
of vector mesons are smaller than $1/3$ in the large $\eta$ region,
but the effect is not  very significant.
In the transversity frame, however,
the $\rho_{00}$'s of 
$K^{*+}$, $\rho^{\pm}$ and $\rho^0$ are larger than $1/3$ and
increase to about 0.5 with increasing
$\eta$, which are similar to the results of 
longitudinally polarized case.
However, the results of $K^{*0}$'s $\rho_{00}$ in both frames are 
much near the unpolarized case of 1/3.
This is because, the $d$ quark fragmentation dominates the $K^{*0}$'s
production at high $p_T$,
and according to the Soffer's inequality, 
$|\delta d(x,Q^2)|$ should be small,
since $\Delta d(x,Q^2)$ is negative for valence quark.

\begin{figure}[t]
\psfig{file=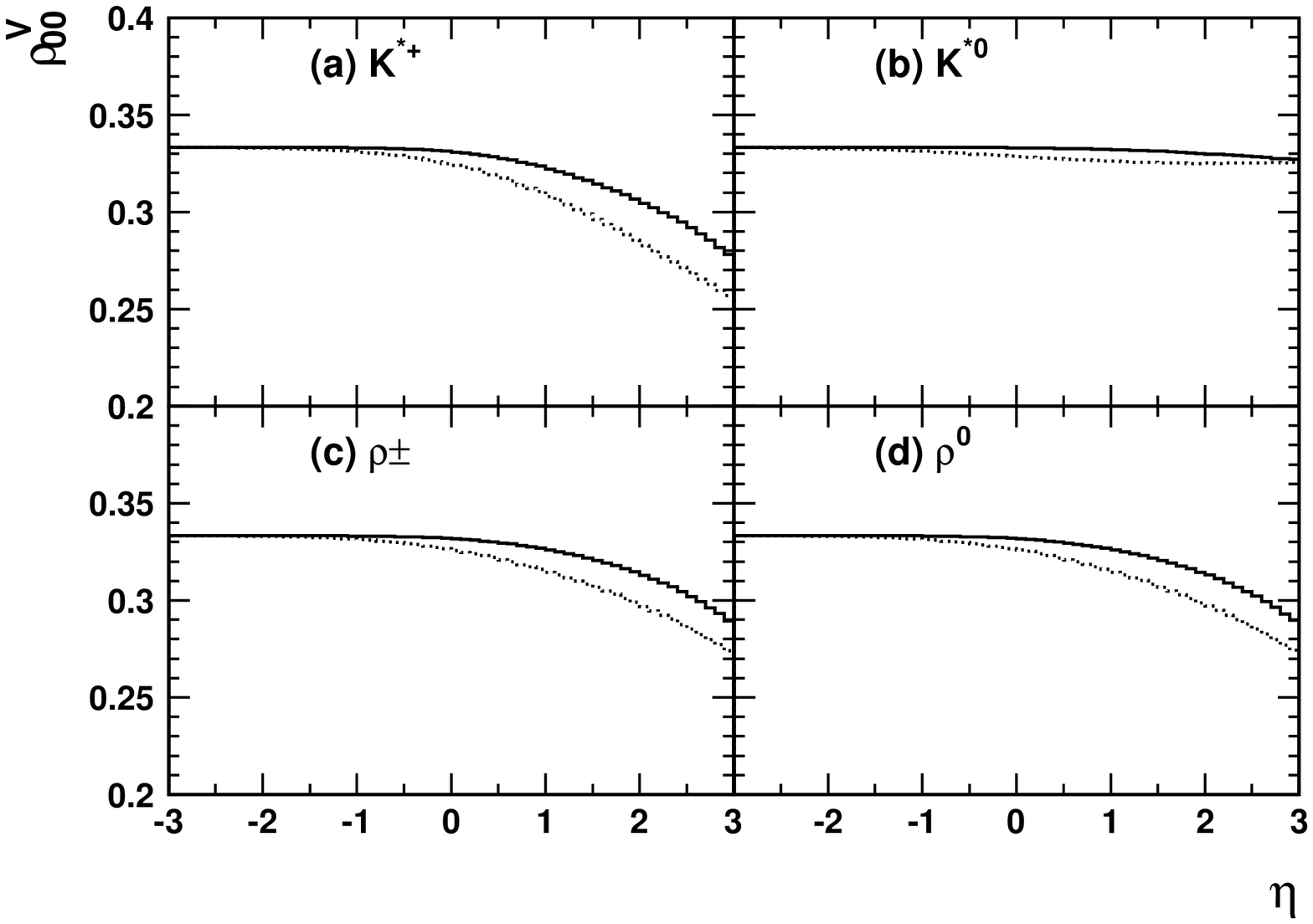,width=15cm}
\caption{The same as Fig. 4, but for $p_T$$>$8 GeV.}
\label{vfigt8h}
\end{figure}

\begin{figure}[t]
\psfig{file=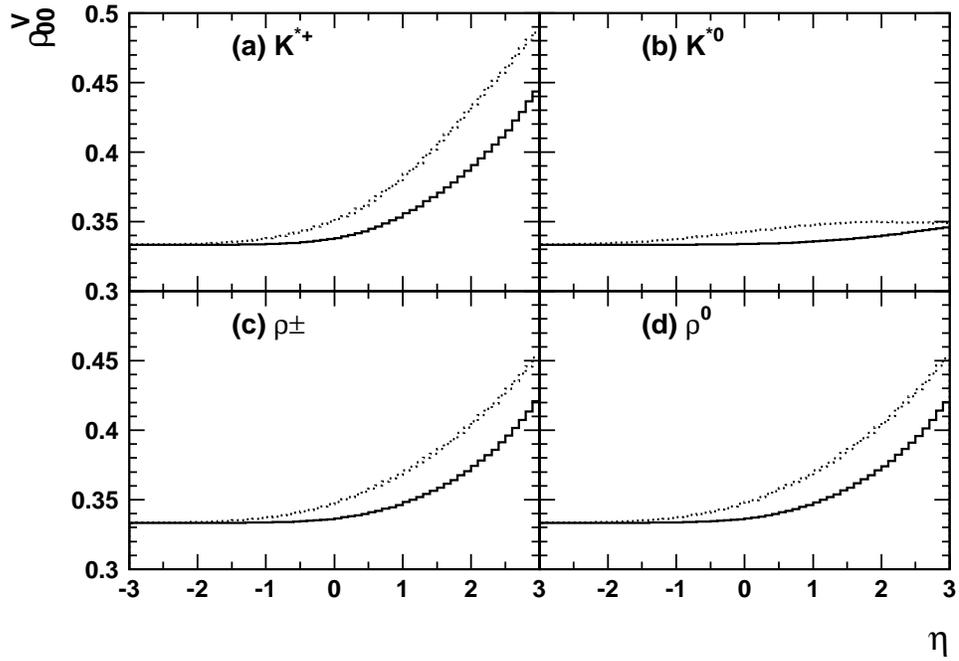,width=15cm}
\caption{The same as Fig. 5, but for $p_T$$>$8 GeV.}
\label{vfigt8}
\end{figure}

As we mentioned above, the gluon is not transversely polarized,
so it is much safer to assume that
there is no contribution of gluon fragmentation
to the spin alignment of vector mesons in transversely polarized 
$pp$ collisions.
Therefore, the high $p_T$ cut which we invoked in longitudinally 
polarized case to ensure a small
contribution of gluon fragmentation is unnecessary now.
Here, only a relative high $p_T$ cut is required to ensure
the validity of pQCD.
We therefore make also the calculations for $p_T>8$ GeV 
at $\sqrt s=500$ GeV and the results are given in 
Fig. \ref{vfigt8h} and Fig. \ref{vfigt8}.
The results are much similar to those obtained 
in the case of higher $p_T$ cut $p_T>13$ GeV.

\section{Summary}
In summary,
we calculate the spin alignment of vector mesons with high $p_T$
in high energy $pp$ collisions with one longitudinally polarized beam 
by taking the spin of vector meson as the sum of the spin of
the scattered quark and that of the secondary
anti-quark produced in the fragmentation process.
The results for different vector mesons at RHIC energy are presented.
These results show that
there exist quite significant spin alignment for 
$K^{*+}$, $\rho^{\pm}$ and $\rho^0$.
We also study the spin alignment of vector mesons
in the case that one of the proton beams is transversely polarized.
Using the same method as that in the longitudinally polarized case,
we obtain the results in the helicity and the
transversity frame.
The results show that, in the large $\eta$ region,
the $\rho_{00}$'s in the helicity frame
are smaller than $1/3$, while the $\rho_{00}$'s
in the transversity frame are larger than $1/3$.
The magnitudes of the effect depend on the transversity
distribution of the quark in proton.
They are expected to be more significant in the large $\eta$ region. 
Measurements of them can provide useful information for spin effects
in strong interaction and the spin structure of nucleon.

\vskip 0.3cm
We thank Li Shi-yuan, Si Zong-guo, Xie Qu-bing
and other members in the theoretical particle physics group of
Shandong University for helpful discussions.
This work was supported in part by the National Science Foundation
of China (NSFC) under the approval number 10175037
and the Education Ministry of China
under Huo Ying-dong Foundation.

\newpage

\newpage

\newpage

\newpage

\newpage

\newpage

\newpage

\newpage

\noindent Figure captions
\vskip 0.3cm 

\noindent 
Fig.1: (a) Contributions of quark and gluon fragmentation
to $\rho^{0}$ production, and (b) the ratio of the quark fragmentation's
contribution to the total rate,
in $p p$$\to$$\rho^{0}X$ at $\sqrt s$=500 GeV and $p_T$$>$13 GeV. 

\vskip 0.3cm 
\noindent
Fig.2: Different contributions to $K^{*+}$ production
in $p p$$\to$$K^{*+}X$ at $\sqrt s$=500 GeV and $p_T$$>$13 GeV
as functions of $\eta$. Here `containing hard u" and similar denote
the contributions of those containing the outgoing quark
from the hard scattering subprocess.

\vskip 0.3cm 
\noindent
Fig.3: Spin alignment of different vector mesons with $p_T$$>$13 GeV
in $p p$$\to$$VX$ at $\sqrt s$=500 GeV in helicity frame
when one proton beam is longitudinally polarized.
The standard LO set of GRSV2000\cite{GRSV2000}
and LO set of GRV98\cite{GRV98} are used for helicity
and unpolarized distribution functions respectively.
The scale $\mu$ is taken as the transverse momentum of the
hard scattering subprocess. 

\vskip 0.3cm 
\noindent
Fig.4: Spin alignment of different
vector mesons with $p_T$$>$13 GeV
in $ p p$$\to$$VX$ at $\sqrt s$=500 GeV in the helicity frame
when one of the proton beams is transversely polarized.
The solid lines denote the results obtained using the light-cone
quark-spectator model for $\delta q(x)$;
the dotted lines correspond to the results obtained using the
upper limit in Soffer's inequality. 

\vskip 0.3cm 
\noindent
Fig.5: Spin alignment of different
vector mesons with $p_T$$>$13 GeV
in $p p$$\to$$VX$ at $\sqrt s$=500 GeV in the transversity frame
when one of the proton beams is transversely polarized.
The solid lines denote the results obtained using the light-cone
quark-spectator model for $\delta q(x)$;
the dotted lines correspond to the results obtained using the 
upper limit in Soffer's inequality.  

\vskip 0.3cm 
\noindent
Fig.6: The same as Fig. 4, but for $p_T$$>$8 GeV.

\vskip 0.3cm 
\noindent
Fig.7: The same as Fig. 5, but for $p_T$$>$8 GeV.


\begin {thebibliography}{99}
\bibitem{Collins93} J. Collins, Nucl. Phys. B{\bf 396}, 161 (1993).
\bibitem{Chen95} K. Chen, G.R. Goldstein, R.L. Jaffe, and X. Ji,
  Nucl. Phys. B{\bf 445}, 380 (1995).
\bibitem{Jaf96} R.L. Jaffe, Phys. Rev. D{\bf 54}, R6581 (1996).
\bibitem{Ada97} A.D. Adamov and G.R. Goldstein,
  Phys. Rev. D{\bf 56}, 7381 (1997); D{\bf 64}, 014021 (2001).
\label{Ada97}
\bibitem{BL98} C. Boros and Liang Zuo-tang,
 Phys. Rev. D{\bf 57}, 4491 (1998).
\bibitem{Kot98} A. Kotzinian, A. Bravar and D. von Harrach,
 Eur. Phys. J. C{\bf 2}, 329 (1998).
\bibitem{Ansel01}M. Anselmino, M. Boglione, U. D'Alesio, and F. Murgia,
 Eur. Phys. J. C{\bf 21}, 501(2001);
 M. Anselmino, D. Boer, U. D'Alesio, and F. Murgia,
 Phys. Rev. D{\bf 63}, 054029 (2001);
 M. Anselmino, M. Bertini, F. Murgia, P. Quintairos,
 Eur. Phys. J. {\bf C11}, 529 (1999).
\label{ABM00}                    
\bibitem{Mas} B.Q. Ma, I. Schmidt and J.J. Yang,
 Phys. Rev. D{\bf 61}, 034017 (2000);
 B.Q. Ma, I. Schmidt, J. Soffer and J.J. Yang,
 Phys. Rev. D{\bf 62}, 114009 (2000);
 Phys. Rev. D{\bf 64}, 014017 (2001),
 erratum-ibid, D{\bf64}, 099901 (2001);
 Nucl. Phys. A{\bf 703}, 346 (2002).
\bibitem{LL2000} Liu Chun-xiu and Liang Zuo-tang, Phys. Rev.
 D{\bf 62}, 094001 (2000);
 Liu Chun-xiu, Xu Qing-hua and Liang Zuo-tang, 
 Phys. Rev. D{\bf 64}, 073004 (2001);
 Liang Zuo-tang and Liu Chun-xiu, Phys. Rev. D{\bf 66}, 075302 (2002).
\label{LXL01}                  
\bibitem{XLL02} Xu Qing-hua, Liu Chun-xiu and Liang Zuo-tang,
 Phys. Rev. D 65, 114008 (2002).
\label{XLL02}  
\bibitem{XLL01} Xu Qing-hua, Liu Chun-xiu and Liang Zuo-tang,
 Phys. Rev. D 63, 111301(R) (2001).
\label{XLL01}
\bibitem{XL02} Xu Qing-hua, and Liang Zuo-tang,
 Phys. Rev. D 66, 017301 (2002).
\label{XL02}   
\bibitem{epr82} I. Cohen {\it et al}., Phys. Rev. D 25, 634, (1982).
\label{epr82}
\bibitem{BEBC87} BEBC WA59 Collab., W. Witttek {\it et al}.,
 Phys. Lett. {\bf B187}, 179 (1987);       
 BBCN Collab., V.G. Zaetz {\it et al}.,
  Z. Phys. {\bf C66}, 583 (1995).
\bibitem{EXCHARM00} EXCHARM Collab., A. N. Aleev {\it et al}.,
 Phys. Lett. {\bf B485}, 334 (2000).
\bibitem{Kress01}
 DELPHI Collab., P. Abreu {\it et al}.,
 Phys. Lett. {\bf B406}, 271 (1997);
 OPAL Collab., K. Ackerstaff {\it et al}.,
 Phys. Lett. {\bf B412}, 210 (1997);
 OPAL Collab., K. Ackerstaff {\it et al}.,
 Z. Phys. {\bf C74}, 437 (1997);
 OPAL Collab., G. Abbiendi {\it et al}.,
 Eur. Phys. J. {\bf C16}, 61 (2000).
\label{Kress01}       
\bibitem{Bigi} I.I. Y. Bigi, Nuovo Cimento {\bf A41}, 581 (1977).
\bibitem{Augustin80} J.E. Augustin and F.M. Renard,
 Nucl. Phys. {\bf B162}, 341 (1980). 
\label{Augustin80}
\bibitem{Donoghue}  J.F. Donoghue, Phys. Rev. {\bf D19}, 2806 (1979). 
\label{Donoghue}
\bibitem{pythia} T. Sj\"ostrand {\it et al}., 
  Computer Physics Commun. 135, 238 (2001).
\bibitem{note1}
  We note that the result of Eq.(5)
  is obtained in the helicity frame of the fragmenting quark.
  But, in experiment,
  the spin alignment of produced vector meson is usually
  measured in the helicity frame of the vector meson. 
  As we mentioned in Ref.[\ref{XLL01}], when the energy of the fragmenting
  quark is high, such as at LEP and RHIC,
  the moving direction of the produced meson is very close to
  that of the fragmenting quark.
  The difference of the final results between these two frames is small.
\bibitem{Trans} J.P. Ralston, D.E. Soper, 
 Nucl. Phys. B{\bf 152}, 109 (1979);
 X. Artru, M. Mekhfi,  Z. Phys. {\bf C45}, 669 (1990);
 R.L. Jaffe, X.D. Ji, Phys. Rev. Lett. 67, 552 (1991);
 Nucl. Phys. B{\bf 375}, 527 (1992);
 V. Barone, A. Drago, P.G. Ratcliffe, Phys. Rep. 359, 1 (2002).  
\bibitem{Collins94} J.C. Collins, S.F. Heppelmann, and G.A. Ladinsky,
 Nucl. Phys. B{\bf 420}, 565 (1994).
\label{Collins94}
\bibitem{BQMa} B.Q. Ma, I. Schmidt, and J. J. Yang,
 Phys. Rev. {\bf D 63}, 037501 (2001).
\label{BQMa}
\bibitem{BQMa98} B.Q. Ma, I. Schmidt, and J. Soffer,
 Phys. Lett. {\bf B 441}, 461 (1998).
\label{BQMa98}
\bibitem{Soffer} J. Soffer, Phys. Rev. Lett. 74, 1292 (1995).
\label{Soffer}
\bibitem{GRSV2000} M. Gl\"uck, E. Reya, M. Stratmann, W. Vogelsang,
 Phys. Rev. D{\bf 63}, 094005 (2001).
\label{GRSV2000}                   
\bibitem{GRV98} M. Gl\"uck, E. Reya, and A. Vogt,
 Eur. Phys. J. {\bf C 5}, 461 (1998).

\end{thebibliography}
\end{document}